# Accurate structural modelling of protein-glycan recognition with GlycoWeave

Huifeng Zhao[1], Jintu Zhang[1], Yifei Xiao[1], Youqiao Qian[1], Silong Zhai[1], Zichang Jin[1], Li Wang[1], Yuchao Wang[2], Hang Dong[2], Yifan Zhao[3], Zhenxing Wu[4], Hongyan Du[1], Yuansheng Huang[1], Bowei Zhao[1], Xiaorui Wang[1, 4], Shukai Gu[1], Jike Wang[1], Zhongjun Li[2, 3], Zhongtang Li[2, 3], Tingjun Hou[1, *], Yu Kang[1, *]

[1]College of Pharmaceutical Sciences, Zhejiang University, Hangzhou 310058, Zhejiang, China

[2]State Key Laboratory of Natural and Biomimetic Drugs, School of Pharmaceutical Sciences, Peking University, Beijing, 100191, China

[3]Ningbo Institute of Marine Medicine, Peking University, Ningbo, 315832, China

[4]Faculty of Synthetic Biology, Shenzhen University of Advanced Technology, Shenzhen, Guangdong, 518107, China

## Corresponding authors

**Yu Kang**

E-mail: yukang@zju.edu.cn

**Tingjun Hou**

E-mail: tingjunhou@zju.edu.cn

## Abstract

Protein–glycan recognition underlies diverse biological processes, yet accurate prediction of protein–glycan complex structures remains challenging owing to the chemical diversity, conformational flexibility and sparse experimental structural coverage of glycans. Here we present GlycoWeave, a glycan-specific diffusion framework that integrates chemically constrained conformational sampling, multiscale molecular representations and multi-source learning. By operating in internal-coordinate space, GlycoWeave preserves glycan connectivity and stereochemistry while efficiently exploring bound conformations. A heterogeneous equivariant network couples atom-level interaction geometry with monosaccharide-level structural context, whereas multi-source, multi-head training integrates experimental complexes with structures refined using classical and quantum mechanics-informed potentials. On a benchmark of 698 protein–glycan complexes, GlycoWeave achieves a 71.9% Top-25 high-quality success rate, compared with 19.2% for AlphaFold 3, while requiring only 0.34 s per sampled pose. GlycoWeave further preserves glycan structural validity, recovers native protein–glycan interactions and reproduces linkage-specific conformational preferences, maintaining high accuracy across diverse glycan architectures and protein environments. These results establish GlycoWeave as an accurate, efficient and chemically faithful framework for modelling protein–glycan recognition. GlycoWeave is freely available at https://cadd.zju.edu.cn/glycoweave.

## Main

Glycans are ubiquitous biomolecules that shape interactions between cells and their environment[1-3]. A large fraction of human proteins are glycosylated, with glycosylation particularly prevalent among cell-surface and secreted proteins[1,4-6]. Protein recognition of these surface-presented glycans, together with free glycans and other glycoconjugates, regulates cell signalling, adhesion, immune responses and host-pathogen interactions[2,7-10]. Altered glycosylation and glycan recognition are associated with many human diseases, including cancer, inflammatory disorders and infection, and increasingly provide opportunities for therapeutic intervention[4,11-16]. Protein-glycan recognition therefore represents a widespread and functionally important mode of molecular recognition in biology[1,17-19].

The structural diversity of glycans, however, presents a distinct challenge for their characterization and prediction[20-23]. Unlike proteins and nucleic acids, glycans are assembled from a broad carbohydrate vocabulary with an intricate and delicate grammar, packing exponentially more information density[1,24-26]. The carbohydrate vocabulary in this work comprises 139 stereochemically defined monosaccharide forms spanning major classes of biologically relevant carbohydrates (**Fig. 1A, Supplementary Figs. 1–3 and Supplementary Note 1.1**). These forms differ in stereochemistry, ring form and chemical modification [1,27-29](**Fig. 1B,C, Supplementary Fig. 2 and Supplementary Note 1.1.2**), and can be connected through multiple linkage positions and anomeric configurations to form linear or branched structures [1,22,27,30-32](**Fig. 1D, top; Supplementary Note 1.2 and Supplementary Fig. 5**). Glycosidic linkages further confer substantial conformational flexibility, notoriously complicating experimental structure determination[21,23,33,34](**Fig. 1D, bottom; Supplementary Note 1.2**). Accordingly, experimental structural coverage remains sparse[21,35-38]. Only 6,715 of 167,616 non-redundant protein sequences (4.0%) in the Research Collaboratory for Structural Bioinformatics (RCSB) are associated with resolved protein-bound glycans, whereas approximately 20% belong to sequence clusters containing at least one glycan-associated protein (**Supplementary Fig. 7**). This highlights a major structural blind spot in glycan-associated structural contexts (**Supplementary Note 1.3**). More critically, the experimental coverage is highly uneven[35,39,40]: available complexes are strongly biased towards short glycans (**Supplementary Fig. 9A,B**), common monosaccharides (**Supplementary Fig. 4**) and recurrent linkages (**Supplementary Fig. 6**), whereas long, branched and modified glycans remain particularly scarce (**Supplementary Fig. 9**). Of 77,562 chemically permitted disaccharide linkage types enumerated across the carbohydrate vocabulary considered here (**Supplementary Fig. 8**), only 312 are observed in the RCSB and only 134 occur at least ten times (**Supplementary Note 1.2.5**). Beyond these gaps in coverage, deposited glycan structures can contain stereochemical misassignments, errors in glycosidic-linkage annotation and implausible ring conformations, complicating their use as structural references[41–43].

Recent advances are expanding the scale and resolution at which glycans can be characterized in biological systems[2,32,44–48]. Large-scale glycomics and glycoproteomics are beginning to map glycan variation across tissues, individuals and disease states[44,49–52], while advances in high-resolution cryo-electron microscopy are beginning to resolve complex glycosylation directly in native macromolecular assemblies[53,54]. Complementary advances in machine learning[55–60] and screening technologies[7,61–63] are broadening glycan identification, structural interpretation and functional interrogation. Together, these advances are making glycans increasingly accessible to systematic analysis and molecular design, creating a growing need for accurate structural models of protein-glycan recognition.

Computational approaches to protein-glycan structure prediction now span carbohydrate-tailored docking methods[64–68] and general biomolecular co-folding models[69–71]. Methods such as Vina-Carb incorporate glycosidic conformational preferences and preserve predefined glycan chemistry[64], but their accuracy depends on explicit exploration and scoring of a rapidly expanding conformational space, which becomes increasingly difficult for long and branched glycans[38,64,72]. Docking-based predictions have also often been supplemented by molecular dynamics (MD) simulations using carbohydrate-tailored force fields, such as GLYCAM06[73]. More recently, AlphaFold 3[69] and Boltz-2[71] have extended co-folding to glycan-containing systems. These models represent carbohydrates as CCD-defined chemical components rather than through glycan-specific representations or conformational constraints[69,71]. Whether such general models can capture the stereochemical and conformational complexity of glycans from sparse experimental data therefore remains unclear[74]. Indeed, a glycan-focused evaluation of AlphaFold 3 (AF3) have reported incorrect anomeric and epimeric configurations, glycosidic-linkage errors and high-energy ring conformations[75] (**Supplementary Figs. 11 and 42**). Broader analyses of co-folding models have identified recurrent violations of carbohydrate stereochemistry, bonding and ring geometry[76]. Accurate prediction of protein-glycan binding modes therefore remains an unresolved challenge, as models must simultaneously preserve glycan chemistry, efficiently explore large conformational spaces and learn transferable recognition patterns from sparse and highly uneven structural data.

Here we present GlycoWeave, a glycan-specific diffusion framework for accurate and chemically faithful prediction of protein-glycan complex structures (**Fig. 1G**). GlycoWeave operates in a chemically constrained internal-coordinate space, updating rigid-body placement and internal torsional degrees of freedom while preserving glycan connectivity, stereochemistry and ring geometry throughout prediction (**Fig. 2D; Methods**). By restricting sampling to the degrees of freedom that govern glycan placement and conformational change, this formulation reduces the effective search space without altering the underlying glycan chemistry[77]. Within this confined conformational space, GlycoWeave jointly represents glycans at the residue and atom levels (**Fig. 1E**), integrating higher-order glycan context with the local interaction geometry underlying protein-glycan recognition (**Fig. 1F**). To broaden learning beyond the sparse experimental structural landscape, a multi-source, multi-head strategy[78,79] combines experimentally determined complexes curated in ProGlyDB, a comprehensive and standardized protein-glycan structural resource assembled in this study (**Methods and Supplementary Note 1.1.7**), with structures optimized by both classical and quantum mechanics-informed force-fields (**Fig. 2E; Methods**). Together, these designs integrate chemically constrained conformational sampling, multiscale molecular recognition and complementary structural supervision within a unified framework for protein-glycan structure prediction.

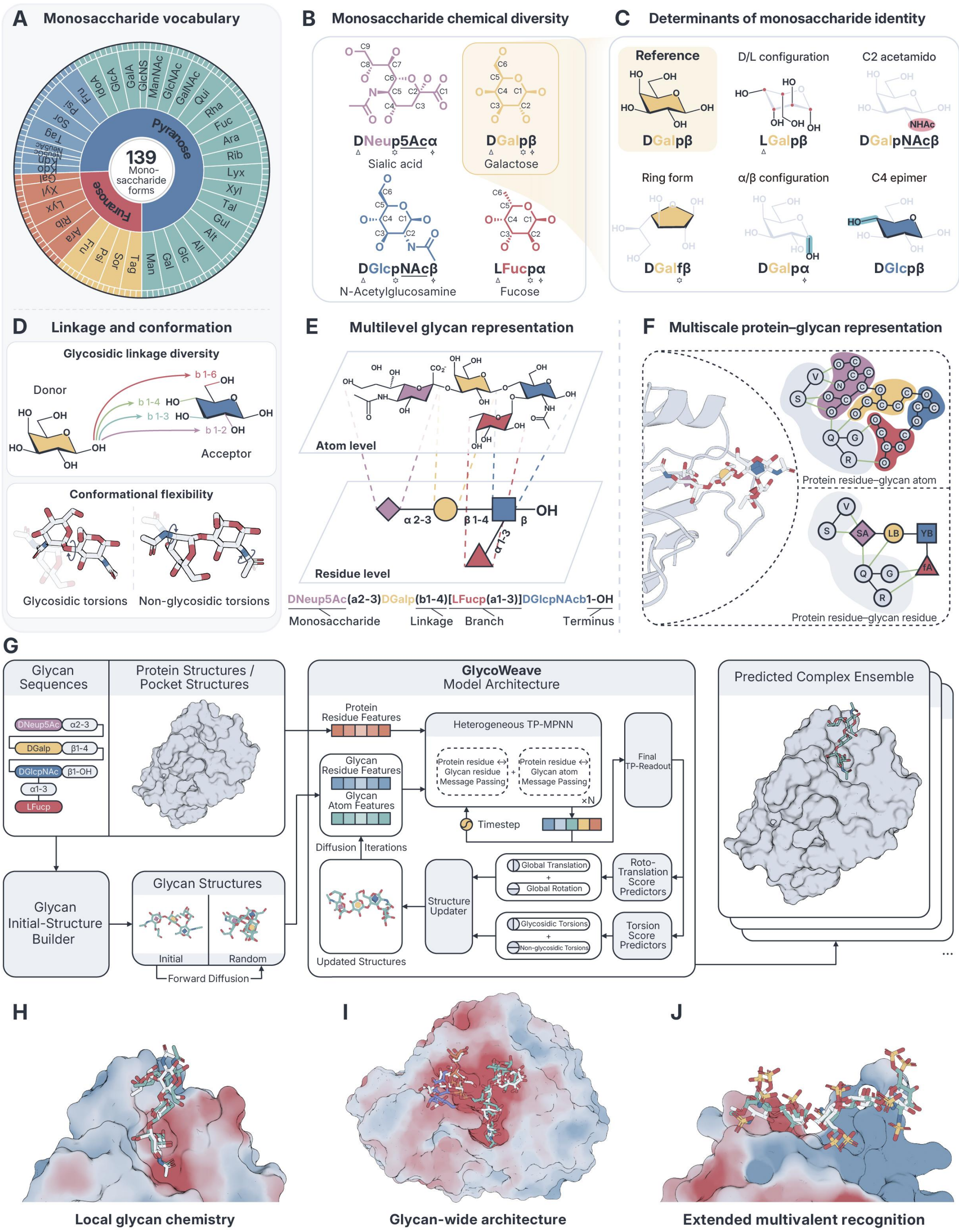

**Fig. 1 | Glycan structural complexity and the GlycoWeave framework for protein–glycan structure prediction.**

**A–D,** Sources of glycan structural complexity, spanning monosaccharide diversity, glycosidic connectivity and conformational flexibility. **A,** Monosaccharide vocabulary comprising the 139 stereochemically defined monosaccharide forms considered in this study. **B,** Chemical structures and condensed labels of four representative monosaccharides: DNeup5Acα (purple), DGalpβ (gold), DGlcpNAcβ (blue) and LFucpα (red), with their common names shown below. In the condensed labels, triangles indicate D/L configuration, stars indicate pyranose/furanose ring form, diamonds indicate α/β anomeric configuration, and underlining identifies substituent or modification descriptors such as 5Ac and NAc. The same monosaccharide-specific colours are retained throughout **B–F** and for the glycan-sequence input in **G. C,** Structural variation relative to the DGalpβ reference, illustrating D/L configuration, C2 acetamido substitution, ring form, α/β anomeric configuration and C4 epimerism. The corresponding notation features introduced in **B** are marked beneath each example; structural features that differ from the reference are highlighted, whereas unchanged atoms and bonds are faded to light grey. **D,** Glycan linkage and conformational diversity. **Top,** alternative glycosidic linkage positions illustrated for a donor–acceptor monosaccharide pair by β1-2, β1-3, β1-4 and β1-6 linkages. **Bottom,** illustrative examples of glycosidic and non-glycosidic torsional degrees of freedom.

**E–G,** Multilevel representation and prediction of protein–glycan complexes in GlycoWeave. **E,** Atom-level, residue-level and IUPAC-condensed sequence representations of a glycan assembled from the four monosaccharides shown in **B.** Glycan residues are represented using Symbol Nomenclature for Glycans (SNFG), and dashed guides connect corresponding residues between the atom- and residue-level representations. Annotations beneath the sequence indicate the monosaccharide, linkage, branch and terminus components of the glycan notation. **F,** Multiscale protein–glycan interaction graphs coupling protein residues to glycan atoms (**top**) or glycan residues (**bottom**). Protein-residue nodes are labelled by one-letter amino-acid codes; glycan-residue nodes contain two-character GLYCAM residue codes together with SNFG symbols identifying the corresponding monosaccharides. Black edges denote covalent connectivity within the protein or glycan, whereas green edges denote distance-based intermolecular protein–glycan interaction edges. **G,** Overview of the GlycoWeave architecture. Glycan sequences and protein or pocket structures are provided as inputs, and the Glycan Initial-Structure Builder generates starting glycan structures that are subjected to forward diffusion. During iterative denoising, the heterogeneous TP-MPNN exchanges information between protein residues and glycan residues or atoms. The final tensor-product readouts feed roto-translation and torsion score predictors, which guide updates of global translation, global rotation, glycosidic torsions and non-glycosidic torsions to generate an ensemble of predicted protein–glycan complexes. **H–J,** Representative protein–glycan complexes illustrating recognition across structural scales. Experimental glycans are shown in white and GlycoWeave predictions in teal; protein molecular surfaces are coloured by APBS-calculated electrostatic potential from negative (red) through near-neutral (white) to positive (blue). **H,** Local glycan chemistry illustrated by the H5N8 haemagglutinin–sialyl-Lewis X complex (PDB 8X2F), highlighting recognition centred on the terminal Neu5Ac residue. **I,** Glycan-wide architecture illustrated by recognition of the highly branched Man9GlcNAc2 glycan by the endo-β-N-acetylglucosaminidase BT1285 (PDB 8U48); the three glycan antennae are additionally colour-coded for visual distinction. **J,** Extended multivalent recognition illustrated by the OVAX–fondaparinux complex (PDB 7QRN), in which a highly sulfated heparin-derived pentasaccharide extends across the protein surface through distributed interactions.

GlycoWeave substantially outperformed state-of-the-art docking and biomolecular co-folding methods on ProGlyDB-RecentSet (**Methods**), achieving an 71.9% Top-25 high-quality success rate (~3.75-fold higher than AF3) while requiring only 0.34 s per sampled pose (~17-fold faster than AF3) (**Fig. 3D and Supplementary Fig. 20**). The framework supports 139 stereochemically defined monosaccharide forms (**Fig. 1A; Supplementary Table 1a**), 4 chemical modifications (**Supplementary Table 1b**) and all their relative theoretically-permitted glycosidic linkages (**Supplementary Fig. 8**), while preserving glycan chemistry and maintaining robust prediction accuracy across diverse structural contexts. Predicted structures also recover native protein-glycan interactions and linkage-specific conformational preferences, showing that gains in pose accuracy are accompanied by physically realistic interaction patterns and glycan conformations. Extensive case studies further show that GlycoWeave captures how protein environments shape glycan conformations and recognition specificity, and resolves binding modes involving local glycan chemistry, glycan-wide architecture and extended multivalent recognition (**Figs. 1H–J and 5A–E**). Together, these results establish GlycoWeave as a rapid, accurate and chemically faithful structural framework for understanding how glycan chemistry, conformational flexibility and protein environments jointly shape molecular recognition.

## A glycan-tailored framework for protein-glycan structure prediction

Accurate protein-glycan structure prediction requires exploration of a large conformational space while preserving the chemical identity of the glycan[80–82]. GlycoWeave therefore performs diffusion in a chemically constrained internal-coordinate space rather than an unrestricted Cartesian space[83,84] **(Fig. 2D and Methods)**. During denoising, the model updates global translation and rotation together with glycosidic and non-glycosidic torsions, subject to constraints on bond geometry, ring closure, stereochemistry, anomeric configuration and modification states. By restricting sampling to the degrees of freedom that govern glycan placement and conformational change, this formulation preserves intended glycan chemistry while reducing the effective conformational search space.

Accurate sampling in internal-coordinate space depends on a chemically appropriate description of local glycan geometry. This geometry is not determined by monosaccharide identity alone, but varies with linkage position, neighbouring monosaccharides, branching state and terminal context[81] **(Fig. 2A)**. We therefore developed a Glycan Initial-Structure Builder that assigns context-specific geometric priors to each monosaccharide, derived from the distributions observed in experimentally resolved glycans **(Fig. 2A–C and Methods)**. In the branched $Man_9GlcNAc_2$ glycan from PDB 8U48, for example, the 11 monosaccharides are mapped into nine distinct local contexts according to their connectivity and neighbouring monosaccharides[85] **(Fig. 2B and Supplementary Note 1.2)**. Even for the same monosaccharide identity, different local contexts can show distinct distributions of bond lengths, bond angles and constrained internal dihedral angles **(Fig. 2C, Supplementary Fig. 5 and Supplementary Note 1.2.4)**. These experimentally derived distributions are then used to assign context-specific geometries when constructing the initial glycan structures for training and inference, providing chemically consistent starting conformations for subsequent sampling.

Within this chemically constrained conformational space, accurate prediction further requires local protein-glycan interactions to be interpreted in the context of higher-order glycan architecture. Binding geometry is defined by local atom interactions at the protein interface, but the contribution of these interactions depends on monosaccharide identity, glycosidic connectivity and position within the broader glycan scaffold[86–88]. GlycoWeave therefore represents each glycan jointly at the residue and atom levels and couples both to protein-residue representations through a heterogeneous tensor-product message-passing network (TP-MPNN)[89,90] (**Fig. 1F,G and Methods**). Information is exchanged across these three representation levels during denoising, allowing each conformational update to integrate local interaction geometry with monosaccharide identity, glycosidic connectivity, higher-order glycan architecture and the surrounding protein environment.

Learning these multiscale recognition patterns, however, is limited by the sparse and uneven sampling of experimentally determined protein-glycan complexes. Although experimental structures provide direct observations of native binding states, they cover only a small fraction of glycan chemical and conformational diversity (**Supplementary Fig. 8 and Supplementary Note 1.2.5**). We therefore expanded structural supervision using 17,006 experimentally determined protein-glycan complexes curated in ProGlyDB-TrainSet (**Methods**), together with approximately 6 million complexes refined by the GLYCAM06j force-field[73] and 4 million complexes optimized by a pretrained Multi-Atomic Cluster Expansion (MACE) potential[91–93] (**Fig. 2E and Methods**). These datasets provide complementary structural information: experimental complexes capture realized protein-glycan recognition modes, GLYCAM06j-refined complexes extend conformational sampling under a carbohydrate-specific classical force-field, and MACE-optimized complexes provide locally refined geometries from a quantum mechanics-informed potential. Because these data share underlying chemical principles but differ in their structural distributions, GlycoWeave shares a heterogeneous TP-MPNN across all three sources together with source-specific score-prediction heads (**Fig. 2E and Methods**). This multi-head strategy allows chemical and geometric information to be shared across data sources while preserving the distinct supervision provided by each structural distribution.

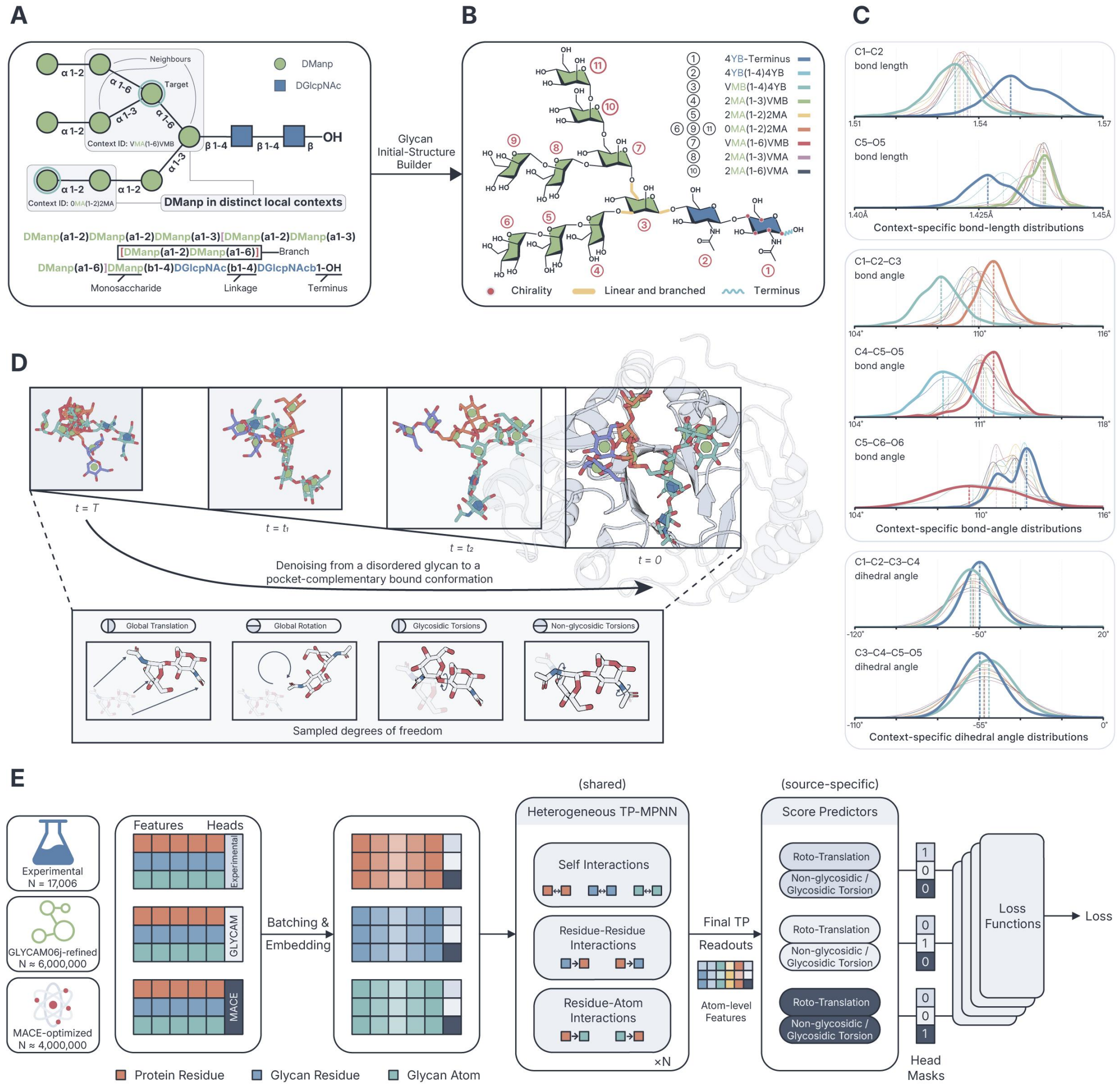

**Fig. 2 | Glycan-tailored design of GlycoWeave.**

**A,** Schematic of the local-context encoding used by the Glycan Initial-Structure Builder. Two DManp residues with the same monosaccharide identity but distinct local environments are used to illustrate context assignment. Target monosaccharides are marked by cyan halos, and their neighbouring monosaccharides and glycosidic linkages define the corresponding context IDs. Monosaccharides are shown using SNFG symbols (DManp, green circles; DGlcpNAc, blue squares). The condensed sequence below annotates monosaccharides, glycosidic linkages, the branch and terminus. **B,** Context assignment for the Man9GlcNAc2 glycan from PDB 8U48. The 11 monosaccharides are numbered by position and mapped to nine local-contexts IDs listed at top right. Matching colors denote the same local context and are retained in **C** for the corresponding geometry distributions. Additional markers denote chirality, linear or branched connectivity and terminal status. **C,** Context-specific distributions of selected bond lengths, bond angles and constrained internal dihedral angles derived from experimentally resolved glycans. Colored dashed lines mark the mode of each kernel density estimate (KDE). For each geometric feature, the two context distributions with the greatest separation between their KDE modes are highlighted with thicker curves. **D,** Schematic denoising trajectory from a disordered glycan at $t = T$ to a pocket-complementary bound conformation at $t = 0$. Sampling is performed over four classes of conformational degrees of freedom: global translation, global rotation, glycosidic torsions and non-glycosidic torsions. **E,** Multi-source, multi-head training architecture. Experimental protein-glycan complexes ($n$ = 17006), GLYCAM06j-refined complexes ($n$≈6 million) and MACE-optimized complexes ($n$≈4 million) are represented at the protein-residue, glycan-residue and glycan-atom levels and processed by a shared heterogeneous TP-MPNN. Within-type message passing and bidirectional protein–glycan message passing operate across the three node types. The updated glycan-atom representations from the shared heterogeneous TP-MPNN are processed by tensor-product (TP) readouts and passed to source-specific score predictors for global translation, rotation, glycosidic torsions and non-glycosidic torsions. Head masks select the corresponding source-specific outputs for loss calculation.

## Accuracy and structural fidelity across protein-glycan complexes

### Chemical and structural validity across glycan space

GlycoWeave generated chemically well-defined protein-glycan complexes for all 698 entries in ProGlyDB-RecentSet. Its supported vocabulary of 139 stereochemically defined monosaccharide forms and four modification types covered every glycan in the set, and all predictions could be analysed directly without post-prediction topology repair (**Supplementary Notes 1 and 2, Supplementary Table 1 and Supplementary Figs. 10 and 11**). We then assessed structural validity at three levels, asking whether the intended glycan identity was retained, whether its internal geometry remained physically plausible and whether the glycan was placed at a reasonable distance from the protein (**Fig. 3A, Supplementary Figs. 12 and Methods**). All 698 Top-1 GlycoWeave predictions passed the complete set of checks. Vina-Carb likewise retained near-complete structural validity (99.7%), whereas only 41.0% of AF3 and 33.7% of Boltz-2 predictions passed the full assessment after the topology standardization required to obtain analysable glycan outputs. Chemical coverage and standardization procedures for the comparison methods are detailed in **Supplementary Note 2**, with the individual validity criteria and failure profiles reported in **Supplementary Note 3.1 and Supplementary Fig. 12**.

## Structural accuracy of protein-glycan complex predictions

With the structural validity of the predicted complexes established, we next assessed how accurately GlycoWeave recovered the experimentally observed binding conformations. Glycan pose quality was evaluated from three complementary aspects: placement of the carbohydrate rings (irRMS), orientation of the glycan scaffold (oRMS) and recovery of native protein-glycan contacts (Fnat), which together define high-, medium- and acceptable-quality predictions (**Methods and Supplementary Note 3.2**). Across ProGlyDB-RecentSet, GlycoWeave achieved a 79.4% Top-1 success rate at the acceptable-quality level, compared with 17.9% for AF3 and 17.8% for Vina-Carb (**Fig. 3B**). The advantage persisted under the most stringent criterion, for which 71.9% of complexes contained a high-quality GlycoWeave prediction within the Top 25, compared with 19.2% for AF3 and 16.9% for Vina-Carb (**Fig. 3D**). The same advantage was maintained at the medium-quality level and across Top-1, Top-5 and Top-25 rankings (**Fig. 3B–D, Supplementary Note 3.2 and Supplementary Fig. 13**). GlycoWeave sampled each pose in an average of 0.34 s under the benchmark conditions (**Supplementary Note 3.5 and Supplementary Fig. 20**).

We further examined prediction quality using CARVE-Score, a continuous metric developed for protein-glycan complex prediction that jointly captures glycan placement, orientation and native-contact recovery. Continuous CARVE-Score comparisons were performed on ProGlyDB-CommonSet, the common evaluable subset across methods, ensuring that distributions and summary statistics were calculated over the same complexes (**Supplementary Note 3.2.1**). Across ProGlyDB-CommonSet, GlycoWeave yielded exceptional performances at the Top-1, Top-5 and Top-25 levels, with its cumulative frequency curves clearly separated from those of the other methods (**Fig. 3G**). AF3 and Boltz-2 retained an advantage over conventional docking methods at very high CARVE-Score cutoffs (>0.9), indicating that they can recover highly accurate structures for a subset of complexes, but this advantage disappeared as the threshold decreased. By contrast, GlycoWeave maintained substantially higher cumulative frequencies over the evaluated score range, indicating broader and more consistent recovery of accurate protein-glycan binding modes across ProGlyDB-CommonSet.

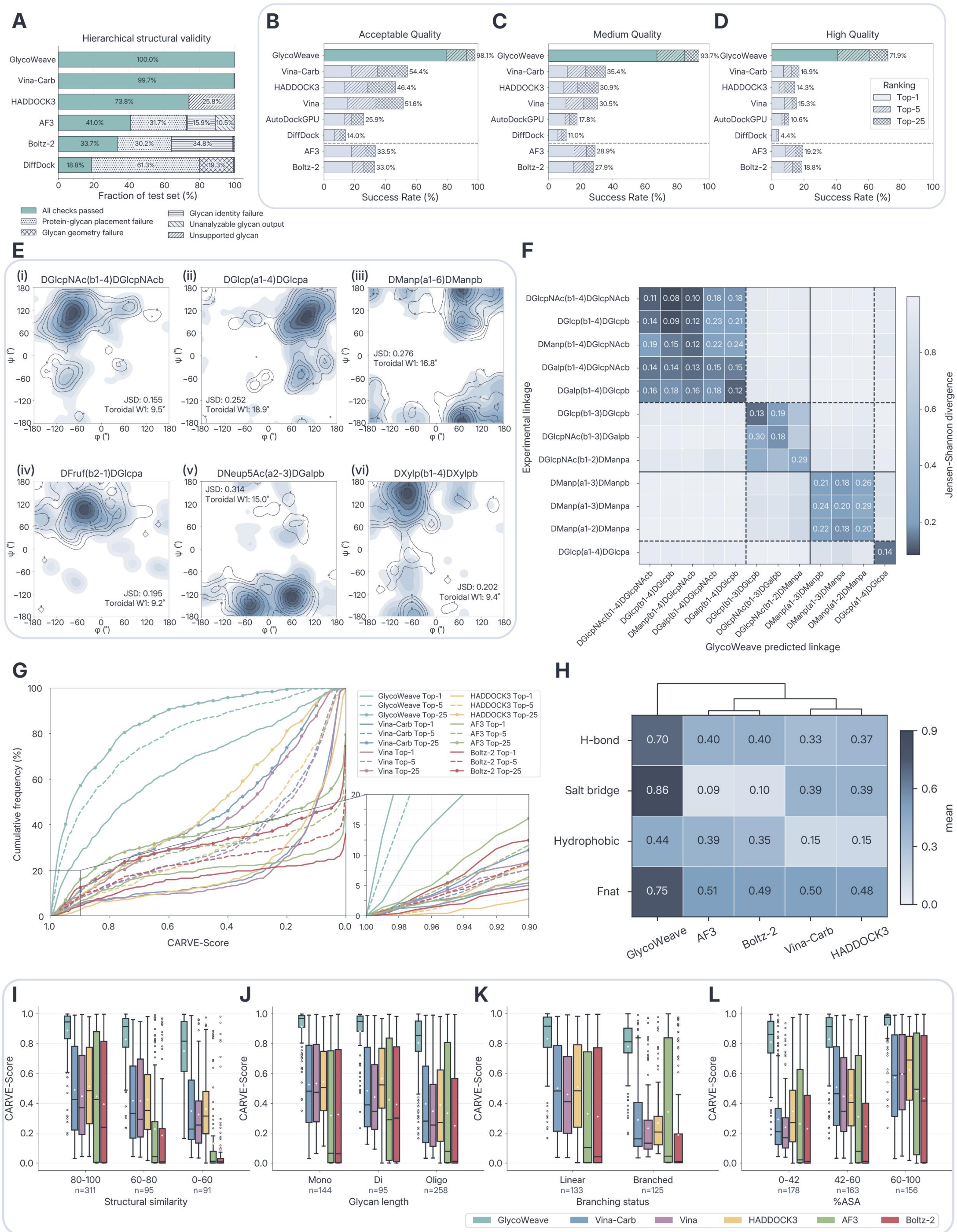

**Fig. 3 | Accuracy and structural fidelity across protein–glycan complexes.**

**A,** Hierarchical structural-validity assessment of Top-1 predictions on ProGlyDB-RecentSet (n = 698), distinguishing predictions that pass all checks from failures in protein–glycan placement, glycan geometry or glycan identity, as well as unanalyzable or unsupported glycan outputs. **B–D,** Top-1, Top-5 and Top-25 success rates for acceptable- (**B**), medium- (**C**) and high-quality (**D**) predictions on ProGlyDB-RecentSet. The dashed line distinguishes methods using a provided protein structure from joint protein–glycan co-folding models. **E,** Comparison of experimentally observed and GlycoWeave-predicted φ–ψ distributions for six representative glycosidic linkages. Numbered contours show the experimental distributions, with lower numbers denoting higher density, and filled density maps show GlycoWeave predictions, with darker shading denoting higher density. Jensen–Shannon divergence (JSD) quantifies differences between probability distributions, whereas toroidal Wasserstein-1 distance (W1) quantifies their separation in periodic torsional space; lower values indicate closer agreement for both measures. **F,** Cross-comparison of experimental and GlycoWeave-predicted φ–ψ distributions for 12 chemically related glycosidic linkages. Heat-map values show pairwise JSD, with darker shading indicating lower JSD. Solid lines separate α- and β-linked groups, whereas dashed lines subdivide finer groups of related linkage types. **G–L**, Analyses on ProGlyDB-CommonSet (n = 497), except **K**, which uses the oligosaccharide subset containing at least three monosaccharides (n = 258). **G,** Cumulative CarbQ distributions for Top-1, Top-5 and Top-25 predictions, shown as the fraction of complexes reaching or exceeding each CarbQ threshold; inset, expanded view for CarbQ > 0.9. **H,** Mean recovery of native hydrogen bonds, salt bridges, hydrophobic contacts and native protein–glycan contacts (Fnat) from Top-1 predictions. Models were hierarchically clustered across the four recovery measures using Euclidean distance and average linkage. **I–L,** CarbQ distributions stratified by maximum structural similarity between test pockets and ProGlyDB-TrainSet (**I**), glycan length (**J**), linear or branched architecture within the oligosaccharide subset (**K**) and binding-site exposure (%ASA; **L**). In **I–L**, boxes span the interquartile range, centre lines denote medians, white points denote means, whiskers extend to 1.5× the interquartile range and points beyond the whiskers denote outliers; n denotes the number of complexes.

## Robustness across structural and glycan complexity

We next examined whether GlycoWeave retained its accuracy across structurally less familiar protein environments and glycans of increasing complexity. Performance declined only modestly as the maximum structural similarity between test pockets and ProGlyDB-TrainSet decreased, with average CARVE-Score remaining 0.75 in the lowest-similarity subset (**Fig. 3I**). By contrast, AF3 and Boltz-2 were considerably more sensitive to this shift, with average CARVE-Score decreasing from 0.45 and 0.39 in the highest-similarity subset to 0.13 and 0.09 in the lowest-similarity subset. Expanding structural supervision beyond experimentally determined complexes markedly improved performance in these low-similarity systems: average CARVE-Score increased from 0.40 with experimental supervision alone to 0.75 after incorporating GLYCAM06j-refined complexes and to 0.78 with the full multi-source, multi-head training regime including MACE-optimized complexes (**Supplementary Note 3.4 and Supplementary Fig. 19**).

This stability extended to glycans of increasing structural complexity. GlycoWeave maintained average CARVE-Score values above 0.8 for monosaccharides, disaccharides and oligosaccharides, whereas the comparison methods showed substantially larger losses for longer glycans (**Fig. 3J**). Among oligosaccharides containing at least three monosaccharides, performance was similarly preserved between linear and branched structures, with average CARVE-Score values of 0.84 and 0.78, respectively (**Fig. 3K**). Binding-site exposure presented a broader challenge, with all methods becoming less accurate for more solvent-exposed glycans. GlycoWeave was much less affected by this change and retained an average CARVE-Score above 0.8 in the most exposed subset with %ASA below 42%, compared with 0.26 for AF3 and 0.29 for Vina-Carb (**Fig. 3L**).

**Recovery of native protein-glycan interactions**

Protein-glycan recognition is mediated by specific networks of non-covalent interactions between the glycan and surrounding protein residues[37]. We therefore asked whether the predicted complexes reproduced the interaction patterns observed in experimental structures. Using the Top-1 predictions on ProGlyDB-CommonSet, we quantified the recovery of native hydrogen bonds, salt bridges and hydrophobic contacts, together with the overall fraction of native protein-glycan contacts (**Fig. 3H and Supplementary Note 4**). GlycoWeave recovered 70% of native hydrogen bonds and 86% of salt bridges, compared with 40% and 9% for AF3 and 33% and 39% for Vina-Carb, respectively. Hydrophobic contacts showed the same overall trend, and GlycoWeave achieved the highest mean Fnat of 0.75, compared with 0.51 for AF3 and 0.50 for Vina-Carb (**Fig.**

**3H)**. Thus, higher pose accuracy was accompanied by more faithful recovery of the intermolecular interactions underlying protein-glycan recognition.

**Recovery of linkage-specific conformational preferences**

Beyond intermolecular recognition, realistic glycan modelling also requires recovery of the linkage-dependent torsional preferences that govern carbohydrate conformation[81]. We therefore compared the glycosidic torsion distributions sampled by GlycoWeave with those observed in experimentally determined structures. Across six representative linkages spanning distinct conformational landscapes, the predicted $\phi - \psi$ distributions reproduced the dominant populations observed in experimental structures and recovered secondary conformational basins where present (**Fig. 3E, Supplementary Note 5.1 and Supplementary Figs. 21–29**). We next examined whether GlycoWeave preserved the conformational specificity of different glycosidic linkages. Cross-comparison of 12 chemically related linkage types showed that, in 9 of 12 cases, each predicted distribution had its lowest Jensen-Shannon divergence from the experimental distribution of the corresponding linkage, while similarities between related linkages followed their shared stereochemical and connectivity features (**Fig. 3F, Supplementary Note 5.2 and Supplementary Fig. 30**). We further examined four linkage types that were absent from the experimental training set. Their experimentally observed conformations fell within the high-probability regions sampled by GlycoWeave (**Supplementary Note 5.3 and Supplementary Figs. 31–34**). These results indicate that the conformational preferences learned by GlycoWeave extend beyond the linkage patterns directly represented by experimental training data.

## Conclusion

Here we have presented GlycoWeave, a glycan-specific diffusion framework that addresses the intertwined challenges of chemical fidelity, conformational complexity and limited experimental structural data in protein–glycan structure prediction. By integrating chemically constrained internal-coordinate diffusion with atom- and monosaccharide-level representations and complementary structural supervision, GlycoWeave enables accurate prediction of protein-bound glycan conformations while preserving their underlying chemical identity.

Across a diverse benchmark of protein–glycan complexes, GlycoWeave substantially outperformed conventional docking and general biomolecular co-folding methods in both structural accuracy and chemical validity, while maintaining high computational efficiency. Its performance remained strong across protein pockets with varying similarity to the training set and glycans of different lengths and branching architectures. Beyond overall binding geometry, GlycoWeave recovered native protein–glycan interaction patterns and reproduced linkage-specific conformational preferences, indicating that its predictions capture important features of both intermolecular recognition and carbohydrate conformational behaviour.

Together, these findings demonstrate the value of incorporating carbohydrate-specific chemical constraints and multiscale structural information into generative molecular modelling. GlycoWeave provides a foundation for investigating protein–glycan recognition across chemically diverse and structurally underrepresented glycan systems.